\documentclass{article}
\usepackage{spconf,amsmath,graphicx,hyperref}
\usepackage{bm}
\usepackage{algorithm}
\usepackage{algorithmic}
\usepackage{booktabs}
\usepackage{capt-of}

\newcommand{\myparagraph}[1]{\vspace{6pt}\noindent\textbf{#1.}\quad}

\title{Source-Directed Trajectory Perturbation at First-Order Cost for Domain Generalization in Speech Deepfake Detection}

\name{Siqing Qin, Kong Aik Lee, Youzhi Tu}
\address{
Dept. of Electrical and Electronic Engineering, The Hong Kong Polytechnic University 
}

\begin{document}
\ninept
\maketitle
\begin{abstract}
Speech deepfake detectors often lose accuracy when the distribution of the test data differs from that of the training data. Meta-learning for domain generalization (MLDG) shows promise by simulating domain shifts with episodic meta-train and meta-test splits. However, the MLDG meta-objective considers only a single clean adaptation trajectory and does not account for the local loss landscape around its endpoints. To address this limitation, we first propose an explicit worst-case MLDG variant, dubbed WC-MLDG-4P, which perturbs both the meta-train and meta-test states but requires four gradient evaluations per episode. We then introduce WC-MLDG-2P, a source-directed alternative to the explicit robust objective. It shifts the clean MLDG endpoint toward a locally higher source-loss state and evaluates the meta-test gradient there, retaining the two-evaluation cost of the first-order MLDG. A first-order expansion relates the resulting gradient change to meta-test directional curvature along the source gradient without explicitly computing a Hessian. Relative to MLDG, WC-MLDG-2P achieves relative mean-EER reductions of 23.4\% with XLSR-AASIST and 7.6\% with XLSR-Conformer-TCM, respectively.
\end{abstract}
\section{Introduction}
\label{sec:intro}
\vspace{-5pt}
Deepfake speech can be produced by many text-to-speech and voice-conversion systems, and new speech generation methods appear faster than detectors can be retrained \cite{asvspoof2019, yamagishi2021asvspoof, asvspoof5, cfad2024, tran2025leveraging}. A detector trained on a fixed set of attacks may therefore perform well on matched data but fail on recordings produced by unseen generators, codecs, or recording conditions \cite{li2025survey,kwok2025bona}. This train--test mismatch is termed domain shift. Domain generalization (DG) studies how to handle such shifts using source-domain data only, without assuming access to the target domain during training \cite{huang2025shift,xie2023dg,ren2025dg}.

A standard approach is empirical risk minimization (ERM), which pools the available source data and minimizes their average training loss. Although ERM is effective under matched conditions, it does not explicitly model how an update learned from one source domain transfers to another. This limitation has motivated meta-learning for domain generalization (MLDG) \cite{li2018mldg}.

MLDG simulates domain shift within each training episode. It divides the source domains into meta-train and meta-test splits, performs a virtual parameter update on the former, and evaluates the updated model on the latter. In speech deepfake detection, attack types are treated as separate domains, and prior MLDG extensions have used multi-task learning, gradient surgery, and parameter-efficient adaptation \cite{qin2026dgsmldg,metatrim2024,hu2022lora,2025metageneralizable}. Their cross-domain gains over ERM suggest that episodic transfer provides a useful training signal.

However, standard MLDG optimizes one clean adaptation path without controlling the local landscape around its starting and adapted states. Either state may still lie in a sharp region, where a small change in the parameters or update direction causes a large loss increase. This matters because unseen generators, codecs, and recording conditions can produce loss geometry different from that of the source domains, motivating a worst-case evaluation of the adaptation trajectory.

Sharpness-aware minimization (SAM) addresses a related problem by perturbing the parameters toward a locally higher-loss state and then minimizing the loss there \cite{foret2020sharpness}. Applied to ERM, it has improved domain generalization in speech deepfake detection \cite{huang25e_interspeech}. Sharp-MAML further incorporates SAM into the lower and upper stages of bilevel MAML for few-shot task adaptation \cite{abbas2022sharpmaml}. Our setting differs in that MLDG uses held-out source-domain splits to simulate domain shift rather than adapting to a new few-shot task. Building on this distinction, we introduce worst-case MLDG (WC-MLDG) for episodic domain transfer. WC-MLDG-4P applies local worst-case evaluation to both ends of the MLDG trajectory. It measures meta-train- and meta-test-side sensitivity directly but requires four gradient evaluations per episode.

We then propose WC-MLDG-2P, a lower-cost source-directed variant. Instead of separately optimizing the two endpoint neighborhoods, it shifts the clean MLDG endpoint along the normalized meta-train gradient and evaluates the meta-test gradient at the shifted state. This construction is distinct from the explicit two-maximization objective, but it retains the two gradient evaluations of first-order MLDG. A Taylor expansion shows that the shifted evaluation measures meta-test directional curvature along the meta-train gradient without explicitly computing a Hessian. With XLSR-AASIST and XLSR-Conformer-TCM, WC-MLDG-2P outperforms ERM and MLDG on all four test sets and remains competitive with WC-MLDG-4P at half the gradient-evaluation count. Our main contributions are summarized as follows:
\begin{itemize}
    \item We instantiate locally robust optimization within MLDG and give an explicit four-pass realization that applies worst-case evaluation to both the meta-train and meta-test states.
    \item We introduce a two-pass source-directed variant that shifts the clean MLDG endpoint without increasing its gradient-evaluation count. We relate this shift to an episode-dependent effective inner step and to meta-test directional curvature along the meta-train gradient.
    \item We evaluate both variants with two detector backends and four unseen test sets, and report their accuracy and cost trade-off.
\end{itemize}

\vspace{-5pt}
\section{Preliminaries}\label{sec:prelim}

\subsection{Meta-Learning for Domain Generalization (MLDG)}
\label{sec:mldg}
Domain generalization seeks to learn parameters $\bm{\theta}$ that perform well on an unseen target domain, using only source domains $\{\mathcal{D}_1,\dots,\mathcal{D}_K\}$, where $K$ is the number of source domains. MLDG \cite{li2018mldg} simulates this setting by randomly splitting them into a meta-train set $\mathcal{S}$ and a meta-test set $\mathcal{T}$. Let $L_{\mathcal S}$ and $L_{\mathcal T}$ denote their average losses. MLDG minimizes the source loss and the target loss after a virtual adaptation:
\begin{align}
  J_{\text{MLDG}}(\bm{\theta})
  &= L_{\mathcal{S}}(\bm{\theta})
  + \beta\,L_{\mathcal{T}}\!\big(\bm{\theta}'_{\text{MLDG}}\big), \label{eq:mldg} \\
  \bm{\theta}'_{\text{MLDG}}
  &= \bm{\theta} - \alpha\, \nabla_{\bm{\theta}} L_{\mathcal{S}}(\bm{\theta}), \label{eq:mldg_inner}
\end{align}
where $\alpha$ is the inner learning rate, $\beta$ is the meta-test weight, and we denote $\bm{g}_{\mathcal{S}} = \nabla_{\bm{\theta}} L_{\mathcal{S}}(\bm{\theta})$. Differentiating $J_{\text{MLDG}}$ with respect to $\bm{\theta}$ requires backpropagating through the inner update, which yields a Hessian--vector product and requires expensive double backpropagation.


To maintain efficiency, first-order MLDG computes $\bm{\theta}'_{\text{MLDG}}$ by stopping gradients through the inner-update direction when differentiating the meta-test loss \cite{finn2017maml}. We denote this operation by $\operatorname{sg}(\cdot)$. It preserves a value in the forward pass and treats it as constant in the backward pass. 
MLDG therefore forms its meta-update along one clean trajectory without incorporating any perturbations, while the local landscapes around the source and adapted states remain outside the objective. 




\subsection{Sharpness-Aware Minimization (SAM)}\label{sec:sam_detail}

Sharpness-aware minimization (SAM) \cite{foret2020sharpness} minimizes the largest loss in a parameter neighborhood of radius $\rho$:
\begin{equation}\label{eq:sam_objective}
  \min_{\bm{\theta}} \max_{\|\bm{\varepsilon}\|_2 \le \rho} L(\bm{\theta} + \bm{\varepsilon}).
\end{equation}
The objective favors parameters whose loss changes little within this neighborhood.

Using a first-order expansion of $L(\bm{\theta}+\bm{\varepsilon})$, the perturbation that increases the local loss most rapidly is aligned with the current gradient:
\begin{equation}\label{eq:epsilon_approx}
    \bm{\varepsilon}^* = \rho \frac{\nabla_{\bm{\theta}} L(\bm{\theta})}{\|\nabla_{\bm{\theta}} L(\bm{\theta})\|_2}.
\end{equation}
SAM then evaluates the training gradient at the perturbed point:
\begin{equation}\label{eq:sam_gradient}
    \nabla_{\bm{\theta}} L_{\text{SAM}}(\bm{\theta}) \approx \nabla_{\bm{\theta}} L(\bm{\theta} + \bm{\varepsilon}^*).
\end{equation}
We apply this idea to both update endpoints of an MLDG episode and then introduce a lower-cost source-directed alternative.

\section{Proposed Method}\label{sec:method}

WC-MLDG applies local worst-case evaluation to the MLDG adaptation cycle. We first present an explicit four-pass objective and then introduce a two-pass source-directed variant. The latter is not algebraically identical to the two-maximization objective and does not separately solve its two local maximizations. Instead, it evaluates the meta-test loss at a source-ascent-shifted MLDG endpoint.

\subsection{Objective of WC-MLDG}
\label{subsec:wcmldg}

The explicit objective evaluates the largest local loss at both ends of the virtual adaptation trajectory:
\begin{equation}
    \min_{\bm{\theta}} J_{\text{WC-MLDG}}(\bm{\theta}) = \underbrace{\max_{\|\bm{\varepsilon}_1\|_2 \le \rho} L_{\mathcal{S}}(\bm{\theta} + \bm{\varepsilon}_\mathcal{S})}_{\text{Worst-case source loss}} + \beta \underbrace{\max_{\|\bm{\varepsilon}_2\|_2 \le \rho} L_{\mathcal{T}}(\bm{\theta}' + \bm{\varepsilon}_\mathcal{T})}_{\text{Worst-case target loss}},
    \label{eq:wc_mldg_objective}
\end{equation}
where the adapted parameters $\bm{\theta}'$ are derived from a standard gradient descent step on the clean source loss:
\begin{equation}
    \bm{\theta}' = \bm{\theta} - \alpha \nabla_{\bm{\theta}} L_{\mathcal{S}}(\bm{\theta}).
\end{equation}
The clean source gradient defines the virtual update $\bm{\theta}\rightarrow\bm{\theta}'$, while the two maximization terms measure local sensitivity before and after that update.

\begin{algorithm}[t]
\caption{Explicit WC-MLDG-4P}
\label{alg:wc_mldg_explicit}
\begin{algorithmic}[1]
\REQUIRE Source-domain pool $\{\mathcal{D}_k\}_{k=1}^{K}$, learning rates $\alpha, \gamma$, weight $\beta$, radius $\rho$.
\WHILE{not converged}
  \STATE $\{\mathcal{S},\mathcal{T}\} \leftarrow \operatorname{RandomSplit}(\{\mathcal{D}_k\}_{k=1}^{K})$
  \STATE $\bm{g}_{\mathcal{S}} \leftarrow \nabla_{\bm{\theta}}L_{\mathcal{S}}(\bm{\theta})$ \COMMENT{Pass 1: clean source}
  \STATE $\bm{\varepsilon}_{\mathcal{S}} \leftarrow \rho\,\operatorname{sg}(\bm{g}_{\mathcal{S}}/\|\bm{g}_{\mathcal{S}}\|_2)$
  \STATE $\bm{g}_{\mathcal{S}}^{\mathrm{WC}} \leftarrow \nabla L_{\mathcal{S}}(\bm{\theta}+\bm{\varepsilon}_{\mathcal{S}})$ \COMMENT{Pass 2: perturbed source}
  \STATE $\bm{\theta}' \leftarrow \bm{\theta}-\alpha\,\operatorname{sg}(\bm{g}_{\mathcal{S}})$ \COMMENT{Clean virtual update}
  \STATE $\bm{g}_{\mathcal{T}} \leftarrow \nabla_{\bm{\theta}'}L_{\mathcal{T}}(\bm{\theta}')$ \COMMENT{Pass 3: clean target}
  \STATE $\bm{\varepsilon}_{\mathcal{T}} \leftarrow \rho\,\operatorname{sg}(\bm{g}_{\mathcal{T}}/\|\bm{g}_{\mathcal{T}}\|_2)$
  \STATE $\bm{g}_{\mathcal{T}}^{\mathrm{WC}} \leftarrow \nabla L_{\mathcal{T}}(\bm{\theta}'+\bm{\varepsilon}_{\mathcal{T}})$ \COMMENT{Pass 4: perturbed target}
  \STATE $\bm{\theta} \leftarrow \bm{\theta}-\gamma(\bm{g}_{\mathcal{S}}^{\mathrm{WC}}+\beta\bm{g}_{\mathcal{T}}^{\mathrm{WC}})$
\ENDWHILE
\end{algorithmic}
\end{algorithm}

\subsection{Explicit Four-Pass Realization}
\label{subsec:explicit_stages}
Algorithm~\ref{alg:wc_mldg_explicit} approximates each inner maximization with the normalized SAM direction and detaches that direction during the following gradient evaluation.

\myparagraph{Source Worst-Case Evaluation}
At the initial state $\bm{\theta}$, the clean meta-train gradient $\bm{g}_{\mathcal S}$ identifies the direction of greatest first-order source-loss increase. It defines the perturbation $\bm{\varepsilon}_{\mathcal S}$, after which the source loss is evaluated again to obtain $\bm{g}_{\mathcal S}^{\mathrm{WC}}$. This gradient accounts for local sensitivity at the starting point of the MLDG trajectory.

\myparagraph{Clean Virtual Adaptation}
The same clean gradient defines $\bm{\theta}'=\bm{\theta}-\alpha\operatorname{sg}(\bm{g}_{\mathcal S})$. Using the clean rather than perturbed gradient preserves the standard MLDG transfer step and keeps adaptation separate from the two endpoint evaluations. The stop-gradient avoids differentiating through the inner-update direction.

\myparagraph{Meta-Test Worst-Case Evaluation and Update}
At the adapted state $\bm{\theta}'$, the clean meta-test gradient defines $\bm{\varepsilon}_{\mathcal T}$, and a second meta-test evaluation gives $\bm{g}_{\mathcal T}^{\mathrm{WC}}$. The final update combines $\bm{g}_{\mathcal S}^{\mathrm{WC}}$ and $\beta\bm{g}_{\mathcal T}^{\mathrm{WC}}$. WC-MLDG-4P therefore measures local sensitivity before and after virtual adaptation, requiring two gradient evaluations on each split.

\subsection{Efficient Source-Directed Trajectory Perturbation}\label{subsec:collapsed_trajectory}

The explicit method directly probes both endpoint neighborhoods but requires four gradient evaluations. WC-MLDG-2P, as shown in Algorithm~\ref{alg:wc_mldg_efficient}, retains a source-directed local test while reusing the clean meta-train gradient to form a shifted MLDG endpoint, so that only one meta-train and one meta-test evaluation are needed.

\begin{algorithm}[t]
\caption{Efficient WC-MLDG-2P}
\label{alg:wc_mldg_efficient}
\begin{algorithmic}[1]
\REQUIRE Source-domain pool $\{\mathcal{D}_k\}_{k=1}^{K}$, learning rates $\alpha, \gamma$, weight $\beta$, radius $\rho$.
\WHILE{not converged}
  \STATE $\{\mathcal{S},\mathcal{T}\} \leftarrow \operatorname{RandomSplit}(\{\mathcal{D}_k\}_{k=1}^{K})$
  \STATE $\bm{g}_{\mathcal{S}} \leftarrow \nabla_{\bm{\theta}}L_{\mathcal{S}}(\bm{\theta})$ \COMMENT{Pass 1: clean source}
  \STATE $\hat{\bm{g}}_{\mathcal{S}} \leftarrow \bm{g}_{\mathcal{S}}/\|\bm{g}_{\mathcal{S}}\|_2$
  \STATE $\bm{\theta}'_{\mathrm{Traj}} \leftarrow \bm{\theta}-\alpha\,\operatorname{sg}(\bm{g}_{\mathcal{S}})+\rho\,\operatorname{sg}(\hat{\bm{g}}_{\mathcal{S}})$ \COMMENT{Source-directed endpoint}
  \STATE $\bm{g}_{\mathcal{T}}^{\mathrm{2P}} \leftarrow \nabla_{\bm{\theta}'_{\mathrm{Traj}}}L_{\mathcal{T}}(\bm{\theta}'_{\mathrm{Traj}})$ \COMMENT{Pass 2: shifted meta-test}
  \STATE $\bm{\theta} \leftarrow \bm{\theta}-\gamma(\bm{g}_{\mathcal{S}}+\beta\bm{g}_{\mathcal{T}}^{\mathrm{2P}})$
\ENDWHILE
\end{algorithmic}
\end{algorithm}

\myparagraph{Two-Pass Construction}
The first pass computes $\bm{g}_{\mathcal S}$. Its negative direction gives the clean virtual adaptation, whereas its normalized positive direction shifts the endpoint toward locally higher source loss. The second pass evaluates the meta-test loss at this shifted state, and the outer update combines the clean source gradient with the shifted meta-test gradient. 

Unlike WC-MLDG-4P, this variant does not separately minimize the worst-case loss at each endpoint. It instead tests how an MLDG update transfers when its endpoint is displaced along a source-sensitive direction. Its shifted endpoint is
\begin{equation}
    \bm{\theta}'_{\text{Traj}} = \bm{\theta} - \alpha\,\operatorname{sg}(\bm{g}_{\mathcal{S}}) + \rho\,\operatorname{sg}(\hat{\bm{g}}_{\mathcal{S}}),
    \label{eq:collapsed_theta}
\end{equation}
where $\hat{\bm{g}}_{\mathcal{S}}=\bm{g}_{\mathcal{S}}/\|\bm{g}_{\mathcal{S}}\|_2$. The first term gives clean virtual adaptation, while the second shifts its endpoint toward higher source loss. Both directions are detached, so meta-test differentiation remains first order. 


\myparagraph{Perturbed-Endpoint}
Equation~\eqref{eq:collapsed_theta} is constructed by first perturbing the original parameters toward a locally adverse source state, $\bm{\theta}_{\mathrm{adv}}=\bm{\theta}+\rho\,\operatorname{sg}(\hat{\bm{g}}_{\mathcal S})$, and then applying the clean MLDG update, $\bm{\theta}'_{\mathrm{Traj}}=\bm{\theta}_{\mathrm{adv}}-\alpha\,\operatorname{sg}(\bm{g}_{\mathcal S})$. The meta-test loss therefore evaluates transfer from a source-sensitive parameter state rather than merely retuning the clean MLDG update.

Because the two displacements are collinear under the first-order construction, the same endpoint also admits the algebraically equivalent form
\begin{equation}
    \bm{\theta}'_{\mathrm{Traj}}
    = \bm{\theta}-\alpha_{\mathrm{eff}}\,\operatorname{sg}(\bm{g}_{\mathcal S}),
    \qquad
    \alpha_{\mathrm{eff}}=\alpha-\frac{\rho}{\|\bm{g}_{\mathcal S}\|_2}.
    \label{eq:effective_step}
\end{equation}
This equality characterizes the endpoint under stop-gradient. The roles of the two coefficients remain distinct: $\alpha$ controls virtual adaptation, whereas $\rho$ controls displacement toward the locally adverse source neighborhood. Consequently, $\alpha_{\mathrm{eff}}$ is not a separately tuned fixed learning rate but an episode-dependent description of the perturbed endpoint. The shift shortens source descent when $\alpha_{\mathrm{eff}}>0$ and becomes source-ascent extrapolation when $\alpha_{\mathrm{eff}}<0$.

\myparagraph{Directional Curvature}
Let $\bm{\theta}'_{\mathrm{clean}}
= \bm{\theta}-\alpha\,\operatorname{sg}(\bm{g}_{\mathcal S})$ denote the clean MLDG endpoint. The efficient WC-MLDG endpoint is
$\bm{\theta}'_{\mathrm{Traj}}
= \bm{\theta}'_{\mathrm{clean}}
+\rho\,\operatorname{sg}(\hat{\bm{g}}_{\mathcal S})$.
A first-order Taylor expansion of the target gradient around
$\bm{\theta}'_{\mathrm{clean}}$ gives
\begin{equation}
\begin{aligned}
\nabla_{\bm{\theta}'}L_{\mathcal T}
(\bm{\theta}'_{\mathrm{Traj}})
\approx\;&
\nabla_{\bm{\theta}'}L_{\mathcal T}
(\bm{\theta}'_{\mathrm{clean}})
\\
&+\rho H_{\mathcal T}
(\bm{\theta}'_{\mathrm{clean}})
\,\operatorname{sg}(\hat{\bm{g}}_{\mathcal S})
+O(\rho^2).
\end{aligned}
\end{equation}
The second term describes the local change in the meta-test gradient along the normalized meta-train direction. It combines the meta-test-loss Hessian with a meta-train update direction and is therefore a cross-domain directional-curvature term.


\section{Experiments}
\label{sec:experiments}
\vspace{-5pt}
\subsection{Experimental Setup}
\label{sec:exp_setup}
\vspace{-5pt}
\myparagraph{Data and domain construction}
All systems are trained only on the ASVspoof 2019 LA (19LA) training set~\cite{asvspoof2019}. Each of its six attack types (A01--A06) forms one source domain, and all bona fide utterances are reused in every domain. We apply RawBoost-4\footnote{\url{https://github.com/TakHemlata/RawBoost-antispoofing}}~\cite{rawboost}, which combines convolutive, impulsive signal-dependent, and stationary signal-independent noise, to each domain. Treating each augmented copy as another domain yields 12 domains, increasing domain diversity and accounting for noise conditions, which are randomly divided into disjoint meta-train and meta-test splits in each episode. ERM-Aug pools the same original and augmented data under one ERM objective. We evaluate on ASVspoof 2021 DF (21DF)~\cite{yamagishi2021asvspoof}, In-the-Wild (ITW)~\cite{muller2022itw}, Fake-or-Real (FoR)~\cite{reimao2019for}, and Codecfake (Codecf)~\cite{Codecfake}; Codecf is reported as the mean EER over its seven subsets. Lower EER is better.

\myparagraph{Detector architectures}
We pair the 300M-parameter XLS-R front-end~\cite{babu2021xls} with either the AASIST graph-attention backend~\cite{tak2022xlsraasist} or the Conformer-TCM backend~\cite{truong24tcm}.

\myparagraph{Training configuration}
Audio is cropped or repeated to 64,600 samples (about 4~s at 16~kHz). We use cross-entropy loss, Adam~\cite{kingma2014adam}, outer learning rate $\gamma=10^{-6}$, weight decay $10^{-4}$, and a batch size of 16. Meta-learning uses $\alpha=10^{-3}$ and $\beta=0.5$. For methods with parameter perturbations, we select the value of $\rho$ that gives the lowest development-set loss. Training stops after seven epochs without a decrease in development loss.
All runs use the same domain pool, random seed, and stopping rule on one NVIDIA RTX 4090 GPU.

We compare ERM, ERM-Aug, MLDG, ERM-Aug-SAM, and the explicit and efficient variants of WC-MLDG. ERM-Aug-SAM applies SAM to the augmented ERM objective. 
This design separates the effects of augmentation, episodic training, and trajectory perturbation, while the two backends test whether the findings depend on a particular detector architecture.

\begin{figure*}[t!]
    \centering
    \includegraphics[width=1.0\linewidth]{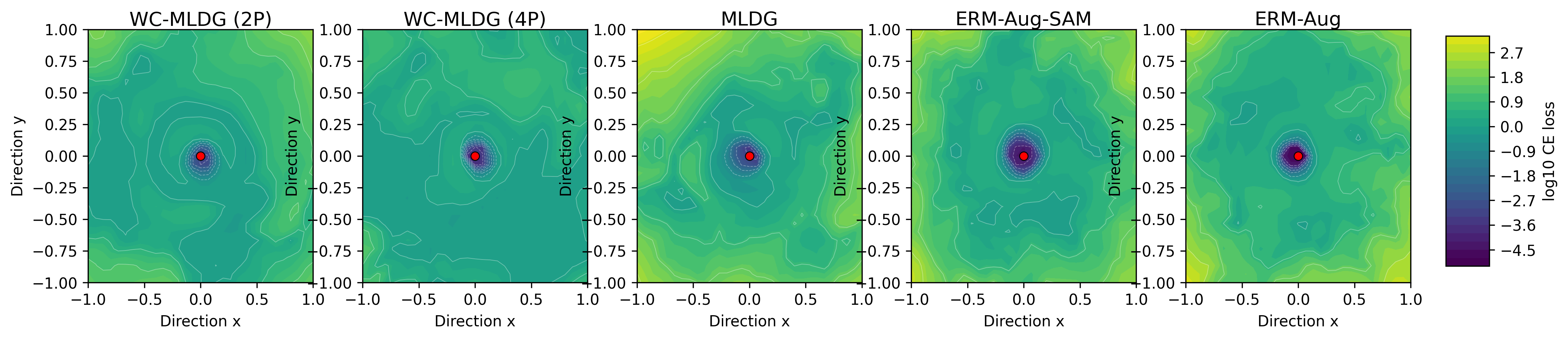}
    \vspace{-2.0em}
    \caption{Local cross-entropy landscapes around the learned checkpoints. Each panel uses the same perturbation range and shared $\log_{10}$ loss scale; the red point marks the unperturbed checkpoint.}
    \label{fig:local_landscapes}
\end{figure*}

\begin{figure*}[t]
\centering
\begin{minipage}[t]{0.69\textwidth}
\vspace{0pt}
\centering
\captionof{table}{EER (\%) on four evaluation sets. ERM-Aug-SAM and WC-MLDG-4P use $\rho=0.001$; WC-MLDG-2P uses $\rho=0.01$. Passes denotes gradient evaluations per iteration. Best results are bold; ``--'' indicates unavailable results.}
\label{tab:main_results_stage1}
\setlength{\tabcolsep}{3.2pt}
\resizebox{\linewidth}{!}{%
\begin{tabular}{llc|cccc|cccc}
\toprule
& & & \multicolumn{4}{c|}{\textbf{XLSR-AASIST}} &
\multicolumn{4}{c}{\textbf{XLSR-Conformer-TCM}} \\
\textbf{Method} & $\bm{\rho}$ & \textbf{Passes} &
\textbf{21DF} & \textbf{ITW} & \textbf{FoR} & \textbf{Codecf} &
\textbf{21DF} & \textbf{ITW} & \textbf{FoR} & \textbf{Codecf} \\
\midrule
ERM              & --    & 1 & 3.69 & 10.46 & 7.47 & 24.41 & 2.06 & 7.79 & 10.69 & 25.43 \\
ERM-Aug         & --    & 1 & 2.61 & 7.21  & 4.90 & 24.83 & 2.11 & 7.97 & 5.31 & 22.86 \\
MLDG             & --    & 2 & 2.69 & 8.43  & 8.48 & 23.96 & 2.19 & 6.47 & 5.91 & 21.79 \\
ERM-Aug-SAM          & .001  & 2 & 2.49 & 6.62  & 3.09 & 26.65 & \textbf{2.00} & 8.84 & 5.48 & 22.66 \\
WC-MLDG-4P     & .001  & 4 & \textbf{1.77} & \textbf{6.18} & 2.96 & 22.28 & \textbf{2.00} & 6.98 & 5.35 & 22.30 \\
\textbf{WC-MLDG-2P} & .01 & 2 & \textbf{1.77} & 7.06 & \textbf{2.91} & \textbf{21.62} & \textbf{2.00} & \textbf{5.96} & \textbf{4.99} & \textbf{20.65} \\
\bottomrule
\end{tabular}
}
\end{minipage}\hfill
\begin{minipage}[t]{0.29\textwidth}
\vspace{0pt}
\centering
\includegraphics[width=\linewidth]{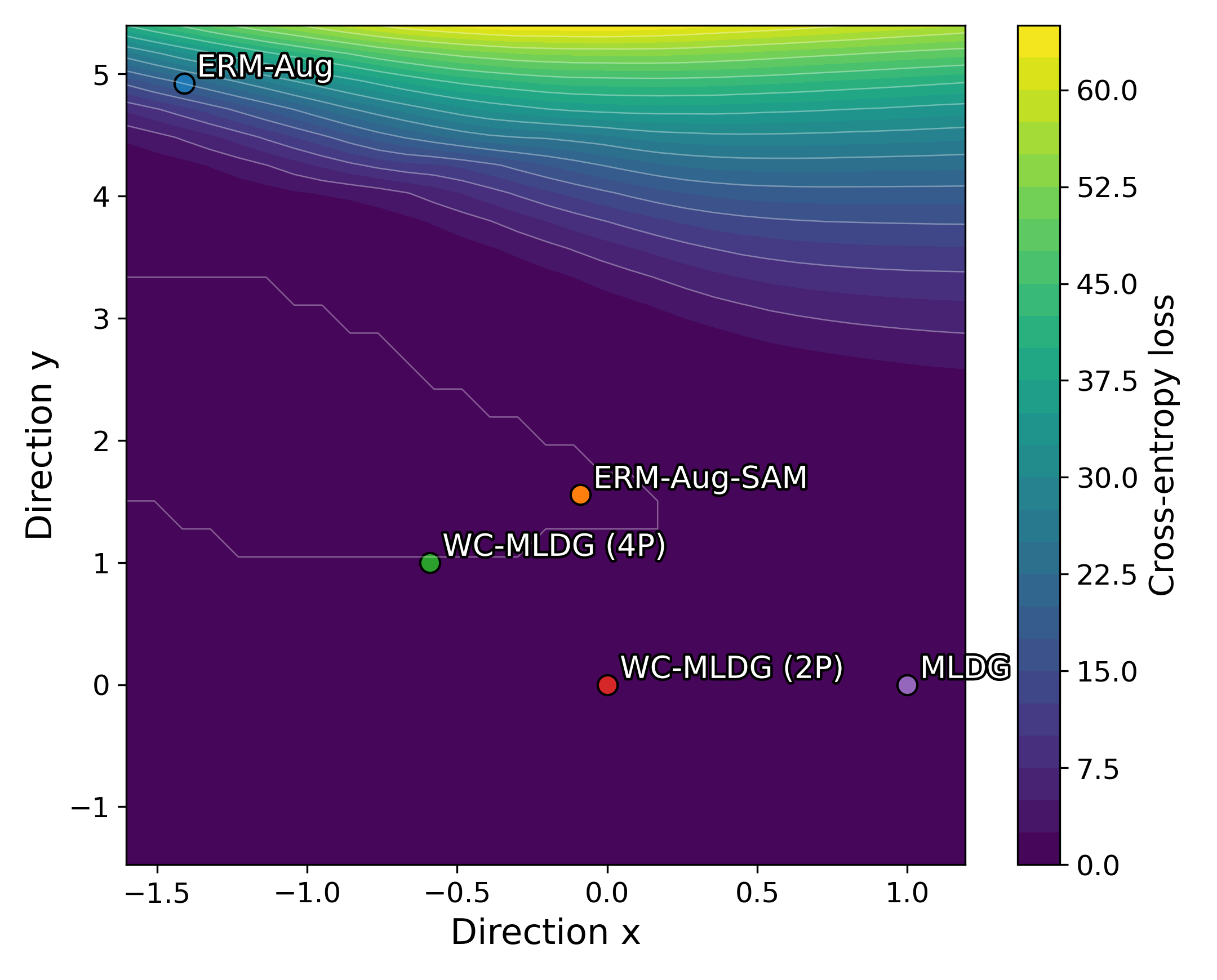}
\vspace{-3em}
\captionof{figure}{Cross-entropy landscape on a common checkpoint plane anchored at efficient WC-MLDG-2P.}
\label{fig:checkpoint_plane}
\end{minipage}
\end{figure*}
\vspace{-3pt}
\subsection{Main Results}
\label{sec:main_results}
\vspace{-3pt}
Table~\ref{tab:main_results_stage1} compares the methods under a fixed configuration. We use one radius for each method on both architectures: $\rho=0.001$ for ERM-Aug-SAM and WC-MLDG-4P, and $\rho=0.01$ for WC-MLDG-2P. We do not select a separate radius for each test set.

\myparagraph{Comparison with ERM-based baselines}
ERM-Aug isolates the effect of extra data augmentation. It improves ERM on three of four AASIST tests and substantially lowers Conformer-TCM EER on FoR, but slightly degrades several other conditions. WC-MLDG-2P improves ERM-Aug in all eight comparisons, including AASIST Codecf (24.83\% to 21.62\%) and Conformer-TCM ITW (7.97\% to 5.96\%). The gains therefore cannot be attributed to augmentation alone.

\myparagraph{Comparison with MLDG}
Standard MLDG does not consistently improve ERM. Its effect changes with the backend and test set. In contrast, WC-MLDG-2P improves MLDG in all eight comparisons. The relative EER reductions are 34.2/16.3/65.7/9.8\% with AASIST and 8.7/7.9/15.6/5.2\% with Conformer-TCM on 21DF, ITW, FoR, and Codecf, respectively.

\myparagraph{Explicit versus efficient WC-MLDG}
Compared with WC-MLDG-4P, the efficient 2P variant ties 21DF on both backends, improves FoR and Codecf on AASIST and ITW, FoR, and Codecf on Conformer-TCM, but is worse on AASIST ITW. It therefore preserves competitive accuracy while halving the gradient-evaluation count.

Overall, efficient WC-MLDG provides the most consistent gains over standard MLDG and ERM-Aug across the two backends, while retaining the two-pass cost of first-order MLDG.

\begin{table}[t]
\centering
\caption{Sensitivity of efficient WC-MLDG-2P to $\rho$ (EER, \%). Dashes indicate settings not evaluated. Bold and underline denote the best and second-best values in each column.}
\label{tab:rho_stage1}
\setlength{\tabcolsep}{5pt}
\begin{tabular}{c|ccc|ccc}
\toprule
& \multicolumn{3}{c|}{\textbf{XLSR-AASIST}} &
\multicolumn{3}{c}{\textbf{XLSR-Conformer-TCM}} \\
$\bm{\rho}$ & \textbf{21DF} & \textbf{ITW} & \textbf{FoR} &
\textbf{21DF} & \textbf{ITW} & \textbf{FoR} \\
\midrule
.1   & -- & -- & -- & \textbf{1.69} & 6.64 & 5.74 \\
.05  & \underline{1.80} & 7.07 & 3.84 & \underline{1.86} & \textbf{5.87} & \underline{5.16} \\
.01  & \textbf{1.77} & \underline{7.06} & \textbf{2.91} & 2.00 & \underline{5.96} & \textbf{4.99} \\
.005 & 2.74 & 8.63 & \underline{3.04} & 2.25 & 6.65 & 5.83 \\
.001 & 1.96 & \textbf{6.09} & 3.17 & 2.43 & 7.58 & 5.70 \\
\bottomrule
\end{tabular}
\end{table}
\vspace{-8pt}
\subsection{Hyperparameter Sensitivity}
\label{sec:rho_sensitivity}
\vspace{-2pt}
Table~\ref{tab:rho_stage1} shows that the effect of $\rho$ depends on the backend and test set. For AASIST, $\rho=0.01$ gives the lowest EER on 21DF and FoR, whereas $\rho=0.001$ performs best on ITW. Conformer-TCM favors larger values on 21DF and ITW, while $\rho=0.01$ gives the best FoR result. No value is best in every column, but $\rho=0.01$ gives the most balanced fixed setting across both backends. A large $\rho$ may place the trajectory endpoint outside the local region described by the Taylor expansion.
\vspace{-5pt}
\subsection{Loss Landscape Visualization}
\label{sec:loss_landscape}
Following the loss-landscape implementation\footnote{\url{https://github.com/tomgoldstein/loss-landscape}}~\cite{visualloss}, we evaluate the 21DF cross-entropy while perturbing all model parameters along two random, filter-normalized directions. In Figure~\ref{fig:local_landscapes}, each checkpoint is placed at the origin and evaluated on the same grid with a shared $\log_{10}$ loss scale. The red point marks the checkpoint; a wider dark region and a slower color change indicate lower local sensitivity. Figure~\ref{fig:checkpoint_plane} instead uses a checkpoint-defined plane anchored at WC-MLDG-2P, with directions toward MLDG and WC-MLDG-4P; the ERM-based checkpoints are shown by their projections onto this plane.

Both WC-MLDG variants have broader low-loss regions and a slower loss increase than MLDG, ERM-Aug-SAM, and ERM-Aug in the visualized planes. Their similar local geometry is consistent with the hypothesis that explicit worst-case evaluation and source-directed endpoint perturbation can lead to less locally sensitive solutions. On the checkpoint-defined plane, WC-MLDG-2P, WC-MLDG-4P, and MLDG are connected through a low-loss region. The projections of ERM-Aug-SAM and ERM-Aug lie progressively farther from this region, with ERM-Aug approaching higher-loss contours. These two-dimensional slices provide qualitative supporting evidence and agree with the EER results in Table~\ref{tab:main_results_stage1}.

\vspace{-0.6em}
\section{Conclusion}
\vspace{-0.4em}
We have introduced the worst-case MLDG (WC-MLDG) for speech deepfake detection. Its explicit realization probes the local loss landscape at both ends of the MLDG trajectory. Its two-pass variant instead evaluates the meta-test gradient at a source-directed perturbation of the clean MLDG endpoint and requires only two gradient evaluations per episode. 
The shifted endpoint admits an episode-dependent effective inner-step, while a local expansion relates its gradient change to meta-test directional curvature. 
Relative to MLDG, WC-MLDG-2P improves all four evaluation sets and achieves relative mean-EER reductions of 23.4\% with XLSR-AASIST and 7.6\% with XLSR-Conformer-TCM.

\vfill\pagebreak

\bibliographystyle{ieeetr}
\bibliography{refs_wc}

\end{document}